\documentclass[preprint]{revtex4}
\usepackage{amsmath}
\usepackage{graphicx}
\usepackage{float}
\usepackage{courier}
\usepackage{hyperref}
\newcommand{\bbsigma}{\boldsymbol{\sigma}}
\newcommand{\bbr}{\boldsymbol{r}}

\begin{document}

\title[]{Structure of the space of folding protein sequences defined by large language models}

\author{A. Zambon$^1$, R. Zecchina$^2$ and G. Tiana$^{1,3}$}
\address{$^1$ Department of Physics and Center for Complexity and Biosystems, Università degli Studi di Milano, Via Celoria 16, 20133 Milano, Italy}
\address{$^2$ Bocconi University, via Roentgen 1, 20136 Milano, Italy}
\address{$^3$ INFN, Sezione di Milano, Via Celoria 16, 20133 Milano, Italy}
\email{riccardo.zecchina@unibocconi.it, guido.tiana@unimi.it}
\vspace{10pt}
\date{\today}

\begin{abstract}
Proteins populate a manifold in the high-dimensional sequence space whose geometrical structure guides their natural evolution. Leveraging recently-developed structure prediction tools based on transformer models, we first examine the protein sequence landscape as defined by the folding score function. This landscape shares characteristics with optimization challenges encountered in machine learning and constraint satisfaction problems. Our analysis reveals that natural proteins predominantly reside in wide, flat minima within this energy landscape.
To investigate further, we employ statistical mechanics algorithms specifically designed to explore regions with high local entropy in relatively flat landscapes. Our findings indicate that these specialized algorithms can identify valleys with higher entropy compared to those found using traditional methods such as Monte Carlo Markov Chains.
In a proof-of-concept case, we find that these highly entropic minima exhibit significant similarities to natural sequences, especially in critical key sites and local entropy. Additionally, evaluations through Molecular Dynamics suggests that the stability of these sequences closely resembles that of natural proteins.
Our tool combines advancements in machine learning and statistical physics, providing new insights into the exploration of sequence landscapes where wide, flat minima coexist alongside a majority of narrower minima.
\end{abstract}

\maketitle

\section{Introduction}

Protein evolution can be described as a stochastic process in the space of sequences. Although it is not possible to predict exactly the course of this process \citep{Lassig2017}, evolution is strongly constrained by functional requirements. One of them is that of foldability, namely that most proteins must display a unique and well-defined native conformation to be functional. This is quite a robust requirement that filters out the vast majority of protein sequences \citep{Shakhnovich1990ImplicationsSequences}.

The space of folding sequences is then a subset of the space of all sequences, whose properties affect the evolution of the protein. Proteins displaying a given function tend to conserve their structure (within an RMSD of 2.5\AA) even among very distant homologous \citep{Sander1991}. Consequently, it is  reasonable to assume that conformational similarity to the wild--type protein is a feature that contributes to the functionality of a mutant, and thus to its evolutionary fitness. Such conformational similarity can be quantified by a distance from the structure of a reference wild--type protein, thus defining a landscape in which evolution is expected to take place through low conformational distance trajectories.

This landscape in sequence space is analogous to the energy landscape of other complex systems studied in physics. In the case of disordered systems, like spin glasses, the energy landscape is rugged \citep{Mezard1984} and its minima are separated by high barriers that prevent diffusion across their conformational space \citep{Mackenzie82}. 
Although the excluded volume of the amino acids is like geometric frustration in glasses and in jamming problems, proteins are quite different from prototypical models of disordered systems. Proteins are small, the hydrophobic amino acids superpose a ferromagnetic-like interaction to the other disordered interactions, and their backbone makes their physical properties quite peculiar.

The space of sequences of proteins that fold to a stable native conformation was studied using minimal protein models that, although not realistic from the biochemical point of view and thus not predictive, display some of the complexity of natural proteins \citep{Govindarajan1997a}. The properties of a sequence in these models are only determined by its native energy because the rest of the conformational spectrum is self--averaging \citep{Shakhnovich1993}; the thermodynamic properties of a sequence are thus determined essentially by the energy $E_N$ of the native conformation. Monte Carlo techniques that control $E_N$ are then a suitable tool for sampling the space of sequences. In this way, it was shown that stable proteins display a complex hierarchical organization with regions not connected by single--point mutations and conserving few mutually interacting residues \citep{Tiana2000}. Nonetheless, it was shown that folding sequences connected by neutral paths can visit vast regions of the space \citep{Govindarajan1997}.

The recent development of machine--learning algorithms \citep{Jumper2021,Lin2022Evolutionary-scaleModel} able to predict the native structure of an input sequence paves the way to studying the space of folding sequences in the context of a realistic, predictive model. Using these algorithms, one can bypass the need of using effective energies, which are not always reliable, to characterize the foldability of a sequence.


We use a fast language model for structure prediction and we combine it with different exploration algorithms. Our method is designed to navigate efficiently through regions of sequence space that have high local entropy (neutral regions). We have put this method to the test on a well studied protein structure, generating predictions that are validated against existing data or through molecular dynamics simulations. The objective of this study is to demonstrate how various innovative approaches, such as language models and algorithms driven by local entropy, can be effectively merged.

The paper is organized as follows: first we describe the methods used to define the effective energy and sample the associated space within the framework of the canonical ensemble. Then, we present the results obtained varying the selective temperature of the system. After selecting a realistic value of the selective temperature, we describe the structure of the energy minima in sequence space, focusing particularly on the width of the corresponding basins. Inspired by the techniques used in connection with artificial intelligence, we finally test an algorithm that can identify large energy minima. We discuss the relevance of these results for protein evolution.

\section{Methods}

\subsection{Sampling the effective energy of a sequence}

In order to sample the space of sequences folding to a given reference conformation $\bbr_0$, we employed a canonical ensemble formalism where each sequence is characterized by an effective energy defined as the fraction of contacts that its native conformation has in common with $\bbr_0$,
\begin{equation}
    E(\bbsigma)=\frac{ \sum_{ij}| \Delta_{ij}(\bbr(\bbsigma))-\Delta_{ij}(\bbr_0) | }{ \sum_{ij}[\Delta_{ij}(\bbr(\bbsigma))+\Delta_{ij}(\bbr_0)] },
    \label{eq:e}
\end{equation}
where $\Delta_{ij}(\bbr)$ ($\Delta_{ij}(\bbr_0)$) is the contact map of the native conformation $\bbr$ ($\bbr_0$), whose elements are 1 if any heavy atom of amino acid $i$ is within 4\AA$\,$ from any heavy atom of amino acid $j$ and 0 otherwise, with $|j-i| > 1$ in order to eliminate the contribution of trivial contacts. Thus, the energy ranges between 0, when all contacts of a sequence are the same as in $\bbr_0$, and 1 if all contacts are different.

The native conformation associated to a generic sequence of amino acids $\bbsigma$ was predicted by ESMFold \citep{Lin2022Evolutionary-scaleModel}, a transformer protein language model defined by approximately 15 billion parameters trained over 65 million protein sequences. The same model was also employed to predict the structure $\bbr_0=\bbr(\bbsigma_0)$ of the reference sequence $\bbsigma_0$.

The sampling was carried out with a Metropolis algorithm \citep{Metropolis1953} at different temperatures $T_s$ (expressed in energy units), that here have the meaning of evolutionary bias towards good (i.e. low--energy) folding sequences. Throughout the simulation, at each step, a random single-site mutation was proposed and the newly generated mutant was accepted or rejected based on its energy, that is the Metropolis rate is here $w(\bbsigma'|\bbsigma)=p_{ap}(\bbsigma'|\bbsigma) \cdot\min[1,\exp(-[E(\bbsigma')-E(\bbsigma)]/T_s)]$, where the {\it a priori} probability is uniform for pairs of sequences with only one different site.

\subsection{Ratcheted sampling} \label{sect:ratchet}

In order to estimate the energy barriers along trajectories from a sequence $\bbsigma_A$ to a sequence $\bbsigma_B$, we employed a Metropolis algorithm which starts from $\bbsigma_A$ and damps the fluctuations in the direction opposite to $\bbsigma_B$. This is based on the principle of the ratchet and the paw and it was used to generate trajectories in the space of protein conformations that resemble physical trajectories \citep{Camilloni2011}.

The Metropolis algorithm is applied with an energy that is given by Eq. (\ref{eq:e}) summed to
\begin{equation}
    E_{r}(\bbsigma(t))=\begin{cases}
        \frac{k}{2}\left[ d(\bbsigma(t),\bbsigma_B)-d_m(t)\right]^2 & \text{if } d(\bbsigma(t),\bbsigma_B) > d_m(t) \\
        0 &  \text{if } d(\bbsigma(t),\bbsigma_B) \leq d_m(t)
    \end{cases}
\end{equation}
where $d_m(t)\equiv\min_{t'<t} d(\bbsigma(t'),\bbsigma_B)$ is the minimum Hamming distance to $\bbsigma_B$ encountered along the trajectory. This time--dependent energy favors the moves towards $\bbsigma_B$, without exerting work to push the system. In this way, the system crosses the lowest energy barriers as in the unbiased trajectories \citep{Tiana2012}.

\subsection{Local Entropy}

The local entropy has been introduced as a tool for analyzing complex energy landscapes in which flat regions coexist with rugged ones, in the context of non--convex neural networks \citep{Baldassi2015}. The local entropy of a discrete system is defined as
\begin{equation}
    S_{T_s,\gamma}(\bbsigma)=\log\left[\sum_{\bbsigma'}e^{-E(\bbsigma')/T_s-\gamma d_{PAM}(\bbsigma,\bbsigma')}\right]
    \label{eq:lentr}
\end{equation}
and is meant to quantify the width of the energy basin around a sequence $\bbsigma$ . Here $\gamma$ is a Lagrange multiplier that controls the average distance from $\bbsigma$.

To define the neighborhood of a sequence, we define a distance
\begin{equation}
    d_{PAM}(\bbsigma,\bbsigma') = N^{-1} \sum_{i} [1 - \frac{P(\sigma_i,\sigma'_i) + P(\sigma'_i,\sigma_i)}{2}]
    \label{eq:dpam}
\end{equation}
that keeps into account the chemical similarity between amino acids, where 
\begin{align*}
    P(\alpha, \beta) = \frac{PAM1[\alpha, \beta]}{\sum_{\gamma \neq \beta} PAM1[\gamma, \beta]} 
\end{align*}
is the transition rate from the $\beta$ to the $\alpha$ amino acid as defined by the PAM1 matrix \citep{Dayhoff1978}, setting the diagonal elements $P(\alpha,\alpha)=1$. A further advantage of $d_{PAM}$ with respect to the Hamming distance $d$ is that it varies essentially as a real variable.

From the identity
\begin{equation}
    \frac{\partial{S_{T_s,\gamma}(\bbsigma)}}{\partial{\gamma}} = -\frac{\sum_{\bbsigma'} d(\bbsigma,\bbsigma') e^{-E(\bbsigma')/T_s-\gamma d_{PAM}(\bbsigma,\bbsigma')}}{\sum_{\bbsigma'}e^{- E(\bbsigma')/T_s-\gamma d_{PAM}(\bbsigma,\bbsigma')}} = -\langle d_{PAM}(\bbsigma,\bbsigma')\rangle_{T_s,\gamma}
\end{equation}
and keeping in mind that
\begin{equation}
    \lim_{\gamma\rightarrow\infty} S_{T_s,\gamma}(\bbsigma) = -E(\bbsigma)/T_s
\end{equation}
one can derive the local entropy difference $\Delta S_{T_s,\gamma}(\bbsigma) \equiv S_{T_s,\gamma}(\bbsigma) - S_{T_s,\infty}(\bbsigma)$ with respect to the single sequence $\bbsigma$. This is found by calculating the integral
\begin{equation}
    \Delta S_{T_s,\gamma}(\bbsigma) = \int_\gamma^\infty \langle d_{PAM}(\bbsigma,\bbsigma')\rangle_{T_s,\gamma'}\,d\gamma',
    \label{eq:ds}
\end{equation}
that can be estimated numerically from simulations performed at different values of $\gamma$.

\subsection{Replica simulations} \label{sect:replicas}

As discussed in ref. \cite{Baldassi2016ent}, from the local entropy measure one can derive several entropy-driven search algorithms. Here we consider Monte Carlo algorithms, as presented in the Methods section, in which it is shown that a replicated MC process focuses on flat dense regions.

Each replica evolved by a Metropolis algorithm based on the coupling potential
\begin{equation}
    E_{rep}(\{\bbsigma_i\}_{i=1}^{y}) = \sum_{i=1}^{y} E(\bbsigma_i) + \gamma^* \sum_{i=1}^{y} \sum_{j \neq i}^{y} d_{PAM}(\bbsigma_i,\bbsigma_j)
    \label{eq:replica}
\end{equation}
where $E(\bbsigma_i)$ is the effective energy of the $i$-th replica (see Eq. \ref{eq:e}) and $\gamma^* = \gamma \, T_s$, with $\gamma$ being the Lagrange multiplier indicated in Eq. (\ref{eq:lentr}). 

In each simulation, we increased slowly that value of $\gamma$, until the $y$ replicas collapsed on a single high-entropy sequence. Eventually, we obtained a single simulation from a simulation. We repeated the whole procedure to collect more sequences.


\section{Results}

\subsection{Thermodynamics of the space of sequences}

\begin{figure}[ht]
    \centering
    \includegraphics[width=\linewidth]{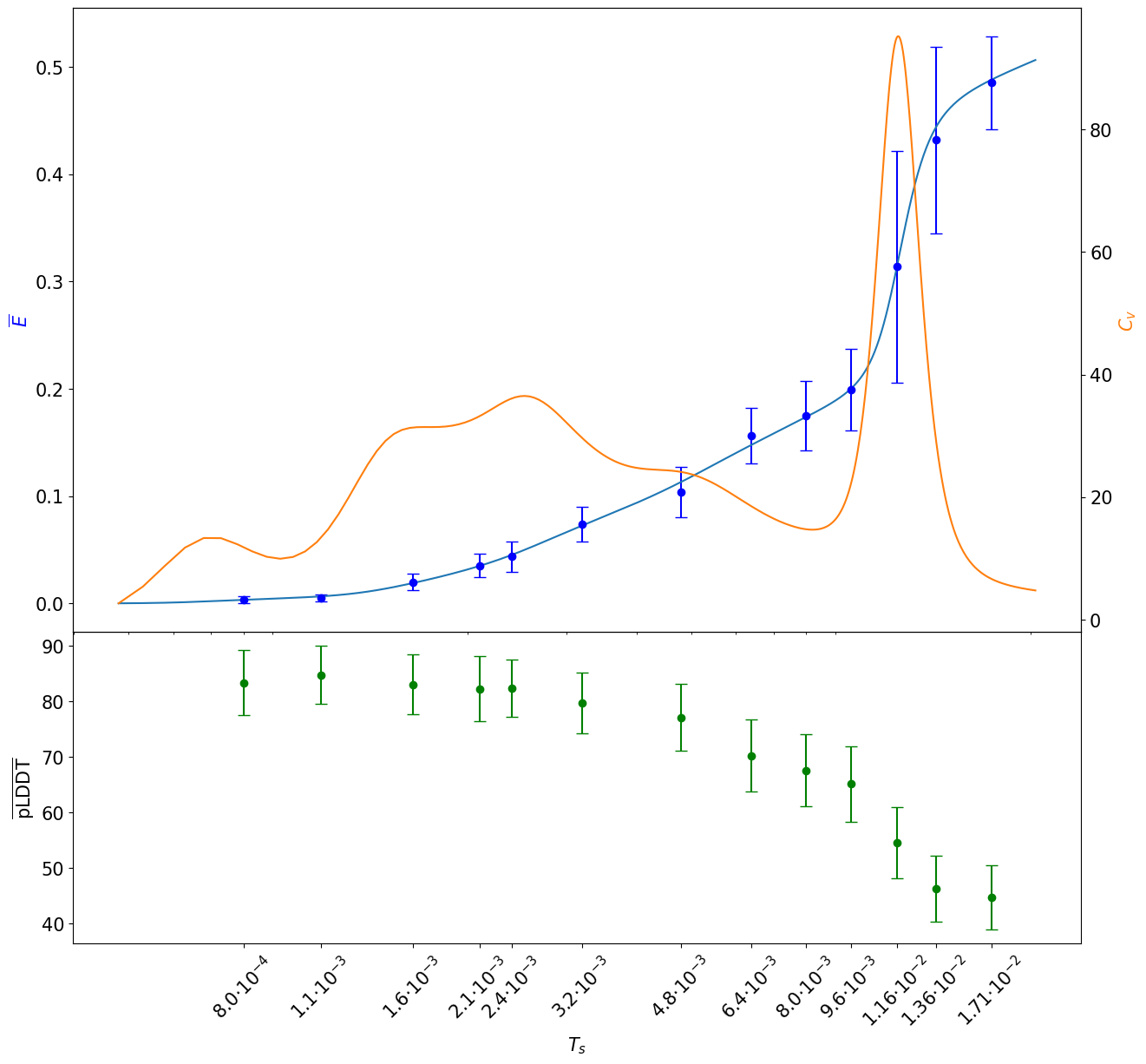}
    \caption{
    Upper panel: the measured average effective energy $\overline{E}$ (blue circles) and the associated specific heat $C_v$ (red circles) calculated from the samplings at different temperatures $T_s$. The solid lines indicate $\overline{E}$ and $C_v$ as estimated from a multiple--histogram algorithm. 
    Lower panel: the values of pLDDT (green circles) representing the average degree of confidence associated to a sequence for its predicted native structure at the different temperatures $T_s$.
    }
    \label{fig:energy}
\end{figure}

We performed samplings of the sequence space at different temperatures $T_s$ for protein G, a widely--studied small protein \citep{McCallister2000} made of an alpha helix and two beta hairpins. Each simulation lasted for at least $\sim 3\cdot 10^5$ steps (see some examples in Fig. S1 and Fig. S2 in the Supp. Mat.); we calculated from them the average energy and the specific heat using a multiple--histogram algorithm \citep{Ferrenberg1989}. The system displays a marked transition at temperature $T^c_s\approx 1.1\cdot 10^{-2}$ between sequences whose native structure has more than 80\% common contacts with the reference structure to sequences with less than 50\% of predicted contacts (upper panel in Fig. \ref{fig:energy}). 
The specific heat also displays a broad shoulder centered at $T_s^n\approx 2.4\cdot 10^{-3}$, at which the average similarity between the contact maps is approximately 95\%.
It should be noted that the typical relative error in the prediction of the contacts of experimentally known structures is approximately $0.1$ (cf. Fig. S4 in the Supp. Mat.), so in the low--temperature phase (i.e. $T_s \leq 3.2\cdot 10^{-3}$), native conformations are indistinguishable from the experimental one. 

It is worth mentioning that, although the space of sequences is combinatorially large, the quantities of interest seem to have reached convergence in the simulation time. To check this, we removed the first steps of the simulation at which the auto--correlation of the Hamming distance from the reference sequence was above $0$, we then divided the rest of the simulation into non--overlapping time blocks and we calculated the average and the standard deviation in each block, showing that they reach stationary values (cf. Fig. S1 and Fig. S2 in the Supp. Mat.).

ESMFold quantifies the degree of confidence in the predicted position of each atom with the pLDDT parameter \citep{Jumper2021}. We calculated the average pLDDT over all the C$_\alpha$ atoms of each sequence sampled at a defined temperature. The average pLDDT (lower panel of Fig. \ref{fig:energy}) of sequences sampled at low temperatures is comparable with that of extant sequences obtained from the pdb, which is 80.1 $\pm$ 12.6 (cf. Fig. S4 in the Supp. Mat.), suggesting that the algorithm is confident that the predicted structures correspond to the unique native state of the protein. The pLDDT is roughly constant at the value of approximately 85 for $T_s<T_s^n$, and then it starts decreasing. However, it remains above 70, which is commonly regarded as the threshold for a good prediction, up to $T^c_s$. At higher temperatures, it drops to 40, which is considered a mark of disorder \citep{Tunyasuvunakool2021}. This suggests that at high temperatures not only sequences are not folding to structures different from the reference one, but they are not folding at all.

Interestingly, at all temperatures, the sequence can step away from the initial, reference sequence (cf. Fig. S3 in the Supp. Mat.). Even at the lowest simulated temperature $T_s=8\cdot 10^{-4}$ at which the effective energy is essentially zero, the average similarity from the initial reference sequence is $\overline{q(\bbsigma, \bbsigma_0)} = 0.37$. This result agrees with the experimental observation that real proteins can change up to $\sim 75\%$ of their sequence while still maintaining their function \citep{Sander1991}.

To have a qualitative validation of the ability of ESMFold to predict correctly the folding properties of sequences that display low energy but are remarkably different than natural ones, we performed some molecular dynamics simulations of three sequences generated by ESMFold (Table \ref{tab:seq}) with a protein model that is regarded as realistically predictive \citep{Robustelli2018a}, starting from the putative native conformation for 200ns at $T=310$K. The three sequences display an average RMSD to the initial conformation of $0.18\pm 0.04$ nm, $0.22\pm 0.06$ nm and $0.25\pm 0.09$ nm, respectively (cf. Fig. S11 in the Sup. Mat). These values are the typical mutual similarities of homologous proteins \citep{Sander1991}, and they are lower than the value $0.36\pm 0.10$ nm found for a selected high--energy sequence.

\subsection{Structure of the space of sequences}

\begin{figure}[H]
    \centering
    \includegraphics[scale=0.5]{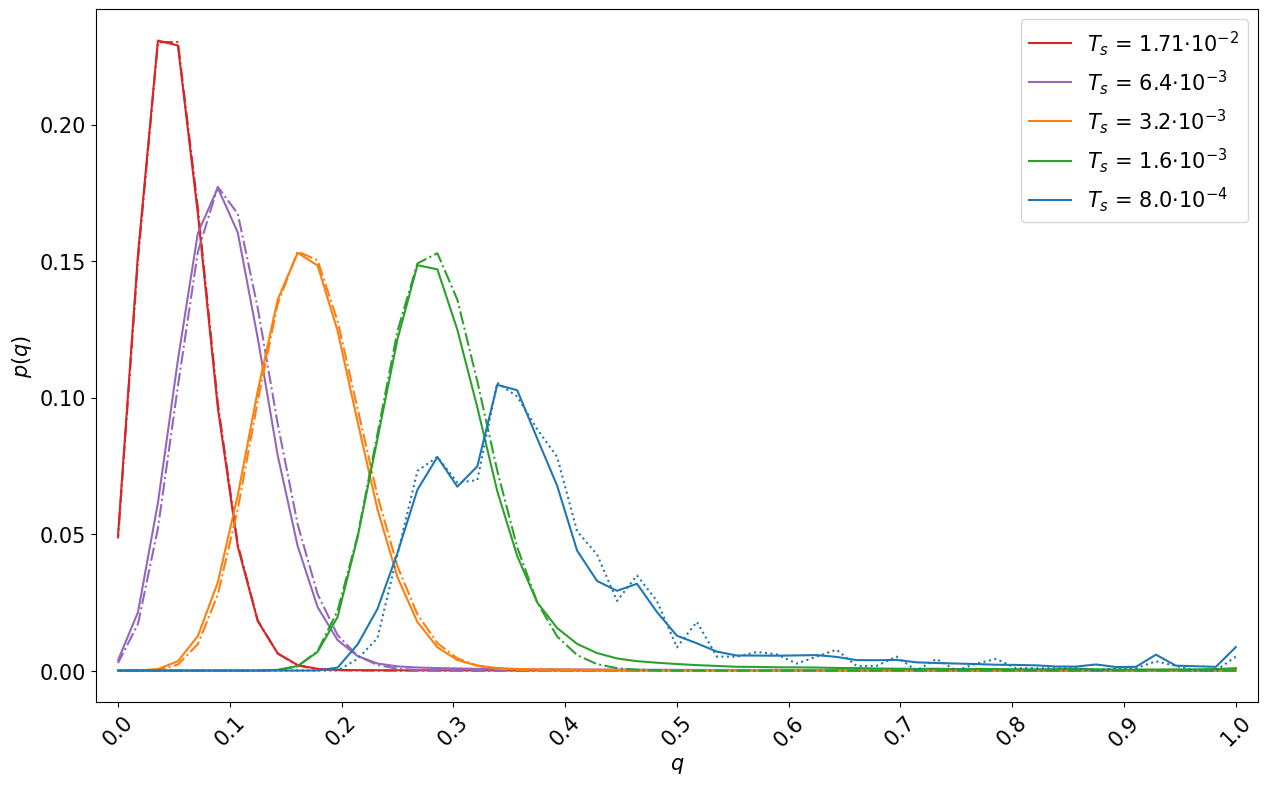}
    \caption{The distribution of sequence similarity $q$ between the pairs of sequences sampled at different temperatures $T_s$. The dot--dashed line is the binomial fit of Eq. \protect\ref{eq:binomial}. The dotted line at $T_s=8\cdot 10^{-4}$ is obtained by decimation of the data, taking one sequence every $10^4$ steps.}
   \label{fig:pq}
\end{figure}

A standard tool used to study the energy landscape of complex systems is the distribution of similarity $q$ between the sampled states \citep{Mezard1984}. In the present case, the value of $q$ between two sequences is defined as the fraction of sites that host the same kind of amino acids. The sampled distribution $p(q)$ displays a unimodal shape in all simulations, whose maximum $q_{EA}$ increases at lower temperatures (Fig. \ref{fig:pq}).

\begin{table}[]
    \centering
    \begin{tabular}{|c|c|c|c|}
    \hline
      $T$                 & $N_f$ & $n_{aa}$  \\
      \hline
      $1.7\cdot 10^{-2}$  & 0     & 20       \\ 
      $6.4\cdot 10^{-3}$  & 0     & 11       \\ 
      $3.2\cdot 10^{-3}$  & 0     & 6        \\ 
      $1.6\cdot 10^{-3}$  & 6     & 5        \\
      $8.0\cdot 10^{-4}$  & 1     & 3        \\
    \hline
    \end{tabular}
    \caption{The results of the fit of $p(q)$ with the model of Eq. \protect\ref{eq:binomial}. The lowest temperature cannot be fitted with a binomial.}
    \label{tab:binomial}
\end{table}

A preliminary interpretation of these curves can be obtained using a very simple model in which the amino acids of the protein of length $N=56$ can vary with uniform probability, except for a number $N_f$ of them that are fixed and identical in all sequences. This gives a binomial distribution
\begin{equation}
    p(q)=\binom{N-N_f}{Nq-N_f}\frac{1}{n_{aa}^{Nq-N_f}}\left(1-\frac{1}{n_{aa}} \right)^{N(1-q)},
    \label{eq:binomial}
\end{equation}
where $n_{aa}$ is the number of different types of amino acids. 

At high temperature ($T = 1.71 \cdot 10^{-2}$) we find $N_f \approx 0$ and $n_{aa} \approx 20$ (see Table \ref{tab:binomial}), with the distribution peak centered at $q_{EA} \approx 1/20$. This is compatible with a state in which amino acids vary essentially at random. Thus, we conclude that for $T_s>T_s^c$ the system displays a single disordered phase.

At lower temperatures (e.g., $T_s=6.4\cdot 10^{-3}$, $T_s=3.2\cdot 10^{-3}$ and $T_s=1.6\cdot 10^{-3}$, Fig. \ref{fig:pq}), the $p(q)$ is still compatible with a binomial distribution. The first two of them (where $T_s^n<T_s<T_s^c$) are fitted with the parameters $N_f=0$ and $n_{aa}<20$ indicating that, according to the minimal model, all residues of the chain can still change, but there is a selection on the type of amino acids they can host.

At the lowest temperature $T_s=8\cdot 10^{-4}$ ($T_s<T_s^n$, at which the predicted structure is essentially identical to the reference one), the shape of $p(q)$ is more irregular, with a tail reaching as maximum similarity $q_M=1$. This suggests that the explored manifold is more complex than at larger temperatures, with energy minima at any mutual distance. The fact that $q_M=1$ indicates that the number of minima is small enough that the probability that the system returns to the same sequences is not negligible. To rule out the possibility that these results are artifacts due to long correlation times in the sampling, we have down--sampled the data, obtaining a distribution almost identical to the original one (dotted line in Fig. \ref{fig:pq}).

To challenge the minimal binomial model, we have then analyzed the degree of conservation of the sites of the protein as a function of the temperature. The site entropy $S(i)\equiv -\sum_\alpha p_i(\alpha)\log p_i(\alpha)$, where $p_i(\alpha)$ is the probability of observing the amino acid of kind $\alpha$ at site $i$, is zero if the site is perfectly conserved and $\log 20\approx 3$ if it displays a uniform probability of hosting the 20 amino acids.\

\begin{figure}[H]
    \centering
    \includegraphics[scale=0.5]{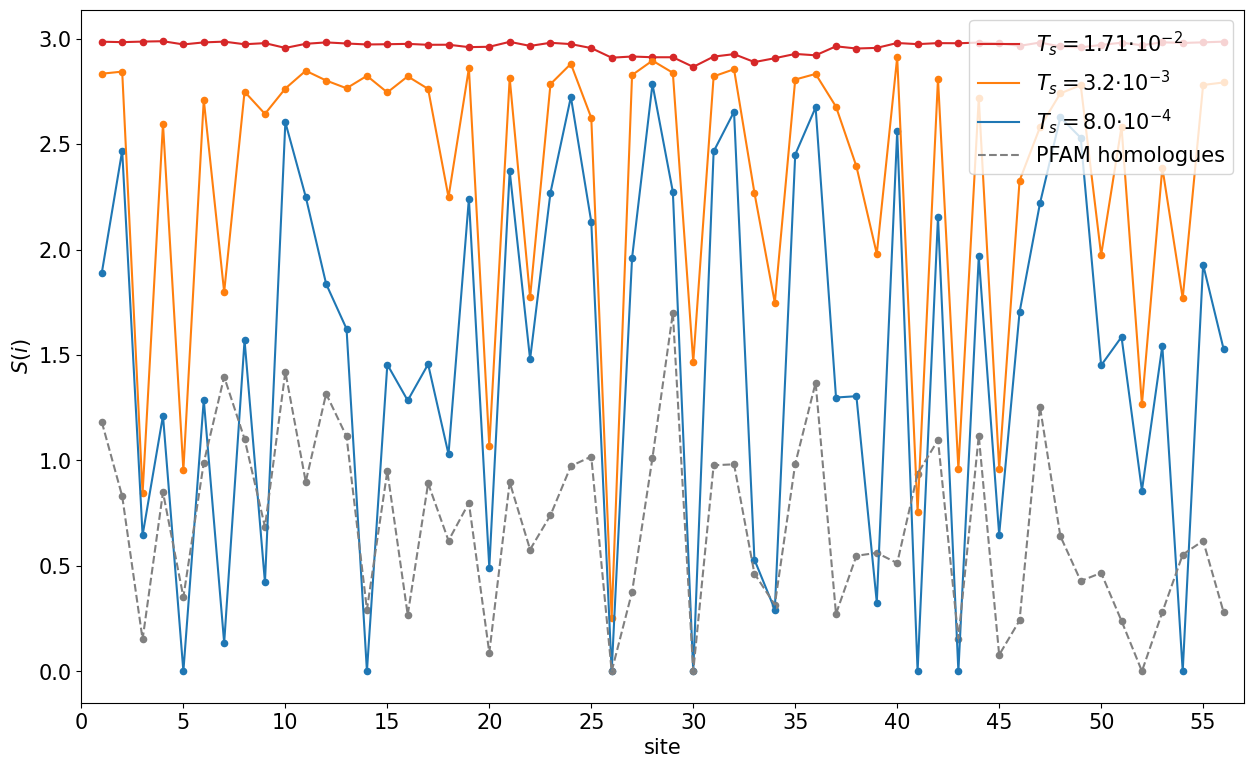}
    \caption{The degree of conservation of the sites quantified by the associated site entropy for the sequences sampled at different temperatures $T_s$ (coloured lines) and for the experimental homologous of protein G (dashed line).}
    \label{fig:siteS}
\end{figure}
\newpage

At high temperatures ($T_s>T_s^c$), the distribution of amino acids is uniform in all sites (Fig. \ref{fig:siteS}).
At the lowest temperature ($T_s<T_s^n$), there are 7 sites that are never mutated and another 9 that are highly conserved, their entropy being lower than 1. Interestingly, approximately one-third of sites display an entropy larger than 2, comparable to that of high--temperature sequences.
At the intermediate temperatures ($T_s^n<T_s<T_s^c$) there is still a (variable) number of low--entropy, highly conserved sites, and a majority of sites whose entropy is comparable to that of high--temperature sequences.

The picture that emerges is that, at all temperatures $T_s<T_s^c$, there is a clear partitioning between highly and poorly conserved sites, and that the main effect of temperature is to define the ratio between the two.

As a consequence, in the case of protein sequences the distribution $p(q)$ (Fig. \ref{fig:pq}) may not contain all the relevant information and it could be misleading, erroneously suggesting that even at low temperatures the system is in a highly disordered state. To overcome this problem, we defined a new distance $d_W$ that weights differently high and low--entropy sites,
\begin{equation}
    d_W(\sigma,\sigma')=\sum_{i\in K} [1-\delta(\sigma_i,\sigma'_i)]+\frac{1}{N-K}\sum_{i\notin K}[1-\delta(\sigma_i,\sigma'_i)],
\end{equation}
where $K$ is the set of sites that display zero entropy at the lowest temperature $T_s = 8\cdot10^{-4}$ in Fig. \ref{fig:siteS}, i.e. 5, 14, 26, 30, 41, 43 and 54. This distance accounts for sequence differences in two disjoint scales, namely units when differences are in $K$--sites and fractions of units when they are not in $K$--sites.

\begin{figure}
    \centering
    \includegraphics[width=\linewidth]{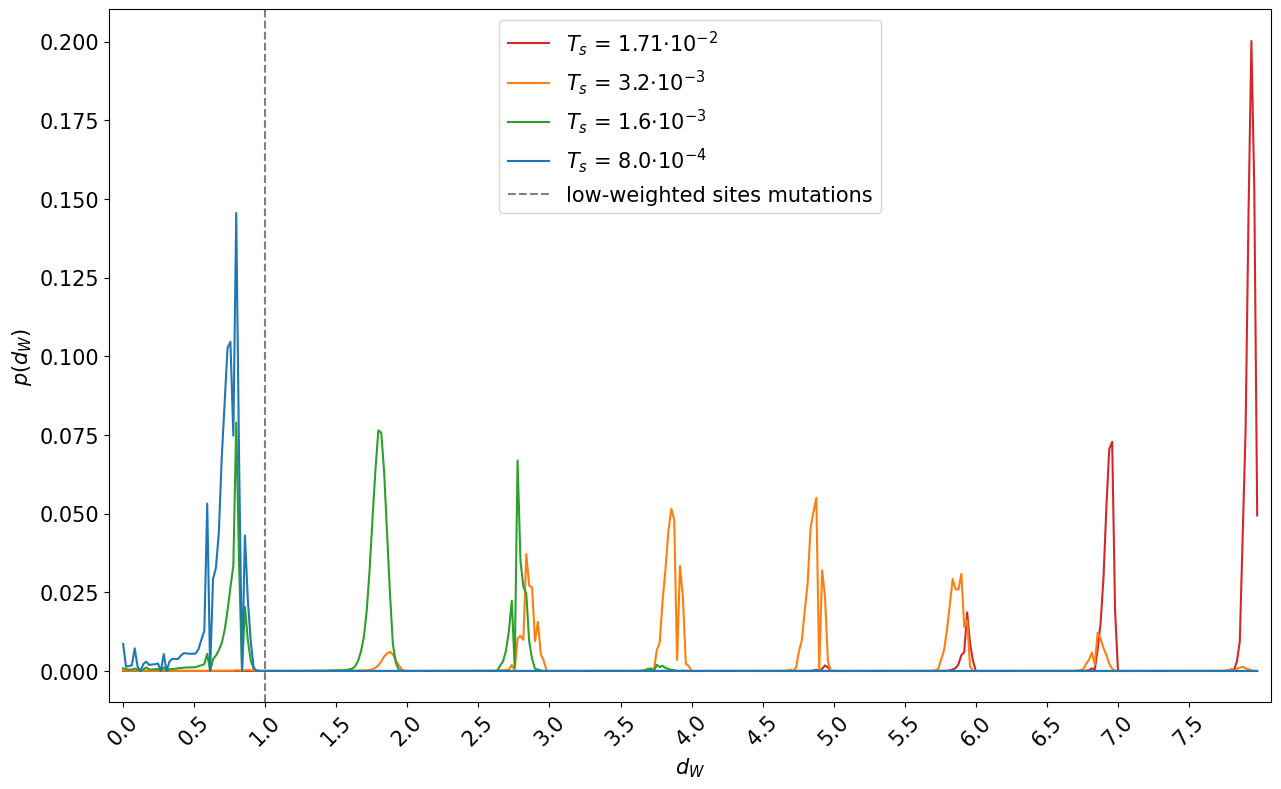}
    \caption{The distribution of distances $d_W$ between the sequences sampled at different temperatures $T_s$.}
    \label{fig:pd}
\end{figure}

At the lowest temperature $T_s=8\cdot 10^{-4}$ ($T_s<T_s^n$), the distribution $p(d_W)$ spans only one unit (Fig. \ref{fig:pd}), since all $K$--sites are conserved by definition. Raising the temperature to $T_s=1.6\cdot 10^{-3}$ ($T_s<T_s^n$), the distribution shifts towards greater values of $d_W$, while still maintaining a peak at $d_W<1$. This indicates that visited combinations of the $K$--sites amino acids are closely distributed in the sequences space, contrary to the information carried by the standard Hamming similarity distribution (cf. Fig \ref{fig:pq}). The distribution of $d_W$ among sequences with the same residues in the $K$--sites ($d_W<1$) has a peak around $0.7$ and a tail to 0, with the same shape as the case $T_s<T_s^n$. This shape suggests that not all the pairs of sequences display the same distance, analogously to the replica symmetry breaking in spin glasses. On the other hand, peaks at $d_W>1$, corresponding to different amino acids in the $K$--sites, are more binomial--like, suggesting that the other sites are uncorrelated. Above $T_s^n$, at $T_s=3.2\cdot 10^{-3}$, the $K$--sites start mutating more freely and the peak at $d_W<1$ vanishes (in agreement with $N_f \approx 0$ found from the binomial model fit, see Table \ref{tab:binomial}), while minor peaks arise at intermediate distances, $d_W\sim2$ and $d_W\sim3$. This is due to the fact that the number of relevant sites for the structural properties of the sequence diminishes as the temperature rises, since the structural constraint on the amino acid sequences becomes less and less rigid. This is also visible in Fig. 3, where at $T_s=3.2\cdot 10^{-3}$, $S(i)>2$ for the $K$--sites $i=14, 30$ and $54$. At $T_s^c$, the $K$--sites vary as all other sites.

\subsection{Comparison with experimental data}

Sequences produced by the natural evolution of protein G can be obtained from the PFAM database \citep{Punta2012}. The main statistical observable that can be calculated from these data and compared with the simulations is the site entropy (dashed curve in Fig. \ref{fig:siteS}).

There are at least two important reasons that make the comparison difficult. First, PFAM sequences are not an unbiased ensemble that reflects the evolution of organisms but they are affected by the choices of researchers to study specific homologous. Moreover, simulations only require that a sequence folds to the correct native state but does not add any functional requirement. This simplification is likely to increase the entropy of sites that lie on the surface of the protein and that are involved in interactions with the cellular environment.

For this reasons, there is no value of $T_s$ at which the entropy of the simulated sequences matches that of the experimental data. Natural sequences conserve non--K--sites much more than any simulation. On the other hand, K--sites are partially conserved similar to what simulations do in the intermediate temperature range.

Studying the Pearson correlation coefficient between the experimental  and the simulated entropy per site (cf. Fig. S6 in the Supp. Mat.), it is clear that there is not a significant difference in correlation for temperatures $T_s < T^c_s\approx 1.1\cdot 10^{-2}$.

In what follows, we shall focus our attention on temperature $T_s=3.2\cdot 10^{-3}$, which belongs to the intermediate regime as experimental data seem to do; at the same time, it is high enough that simulations are computationally fast.

\subsection{Ruggedness of the landscape is mainly determined by changes in K--sites }

An interesting feature of the energy landscape of sequences is its ruggedness. To investigate this point, we generated some artificial low--energy sequences, starting from random ones and quenching the temperature (see Table \ref{tab:seq}), studying the energy landscape along the trajectories that link them to the protein G sequence (1PGB in Table \ref{tab:seq}).

We used a ratchet algorithm (see Methods, Sect. \ref{sect:ratchet}) to generate trajectories at $T_s=3.2\cdot 10^{-3}$ between pairs of sequences. This algorithm does not push the sequence toward its target but only dumps fluctuations in the opposite direction. Consequently, we expect that it will not force the system to cross barriers higher than those that would cross spontaneously by thermal fluctuations \citep{Tiana2012}. 

\begin{figure}
    \centering
    \includegraphics[width=\linewidth]{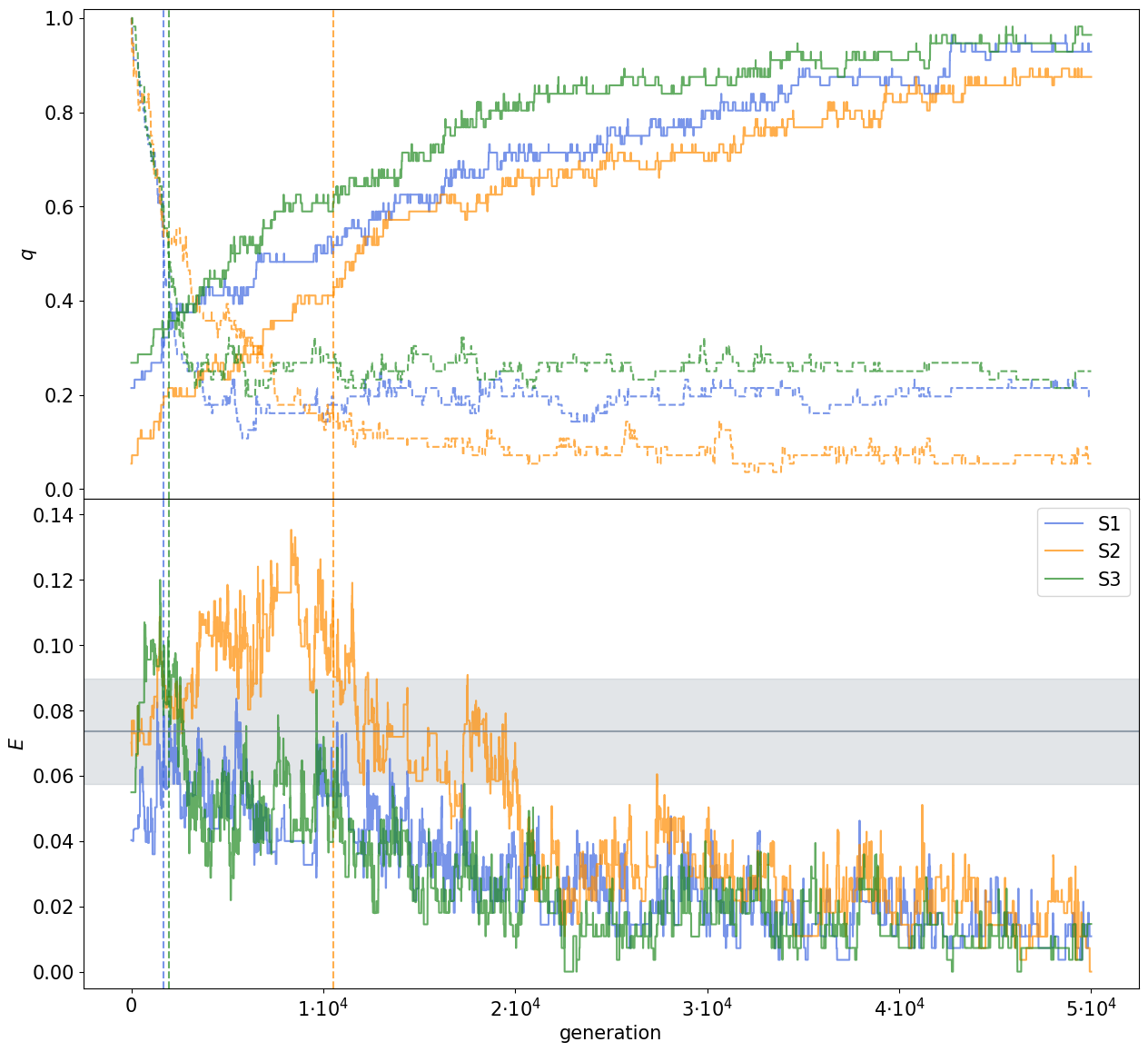}
    \caption{The Hamming similarity parameter $q$ (upper panel) and the effective energy $E$ (lower panel) along trajectories generated with a ratchet algorithm between the system and the 1PGB sequence at $T_s=3.2\cdot 10^{-3}$. The Hamming similarity parameter $q$ is calculated from the initial (dashed curve) and from the target 1PGB (solid curve) sequence along each trajectory. The grey line and the shaded area in the lower panel are the mean energy and the associated standard deviation, respectively, at $T_s=3.2\cdot 10^{-3}$. The vertical dotted lines mark the time when the K--sites approache the target combination of the 1PGB sequence. In the legend, the starting sequences for each trajectory (cf. Table \ref{tab:seq}).}
    \label{fig:ratchet}
\end{figure}

Trajectories can leave the initial sequence in a few thousand mutations and reach the target sequence in less than $10^5$ mutations (upper panel in Fig. \ref{fig:ratchet}). 

The maximum energy reached by the simulation is in the range between $0.09$ and $0.14$ (lower panel in Fig. \ref{fig:ratchet}), which is larger than the spontaneous fluctuations that the system displays at this temperature, at which the mean effective energy is $\overline{E} = 0.074\pm 0.016$ (cf. Fig. \ref{fig:energy}). This fact indicates that the system can encounter relevant energy barriers along its motion. The peak in the energy is close to the time when the K--sites approach that of the target sequence (dashed vertical lines) (cf. Fig. S8 in the Supp. Mat.). The peak is largest for sequence S2, which displays the most different K--sites combination from that of the protein G (cf. Tab. \ref{tab:seq}).

After the K--sites are changed, all sequences take several tens of thousands generations to reach the target sequence. However, mutations of amino acids in sites that are not K--sites do not generate energy barriers comparable to thermal fluctuations.

Summing up, trajectories between low--energy sequences are neutral except for changes in K--sites, which generate barriers that are anyway surmountable by thermal fluctuations.

\begin{table}[]
    \centering
    \centerline{
        \begin{tabular}{|c|c|l|}
        \hline
            id & $E$ & sequence \\
        \hline
           1PGB  & 0     & {\texttt{\small MTYK{\textbf L}ILNGKTLK{\textbf G}ETTTEAVDAAT{\textbf A}EKV{\textbf F}KQYANDNGVD{\textbf G}E{\textbf W}TYDDATKTFT{\textbf V}TE }}\\
           E1NUT2 & 0.17  & {\texttt{\small TVYH{\textbf F}QYDKKGTS{\textbf I}RQDFAAVNKEI{\textbf A}EMH{\textbf F}KEYATESGLD{\textbf A}H{\textbf F}AYNEANQTFV{\textbf Y}KD}}\\
           K9EXL1A & 0.31  & {\texttt{\small EVYT{\textbf F}YYRTQNQN{\textbf G}ATTVKASSPRE{\textbf A}LEY{\textbf F}QNFLSERGLD{\textbf F}N{\textbf W}HYESEDRVFT{\textbf A}SE}}\\
           K9EXL1B & 0.14  & {\texttt{\small AVYT{\textbf F}VYNTKGKN{\textbf G}ATTVKASSPEE{\textbf A}LEY{\textbf F}QNWAKENDLE{\textbf L}D{\textbf W}SYDEDTKTFT{\textbf G}RE}}\\
           S1    & 0.04  & {\texttt{\small PTYR{\textbf M}EVMSTHFE{\textbf A}VVGIEAPNYPA{\textbf A}LHG{\textbf F}VLFCHCLGVL{\textbf A}Q{\textbf F}TYCATHNFFK{\textbf V}WQ}}\\
           S2    & 0.07  & {\texttt{\small HWYR{\textbf F}VHHGPNHE{\textbf C}MGVARVPHVHW{\textbf L}MNA{\textbf V}EKATKAANIK{\textbf C}K{\textbf Y}RWSARHRTLW{\textbf C}YT}}\\
           S3    & 0.06  & {\texttt{\small HEYS{\textbf C}MLISPLRT{\textbf A}TQVFEATNRAM{\textbf A}HWF{\textbf F}EDMALWLGYI{\textbf K}K{\textbf W}TYNERFHMYT{\textbf V}TF}}\\
        \hline
        \end{tabular}
    }
    \caption{Some of the sequences used in the calculations. 1PGB, E1NUT2, K9EXL1A and K9EXL1B are natural sequences labeled by their pdb code. S1, S2 and S3 are artificial sequences obtained by quenching the temperatures in Monte Carlo samplings. In bold, the K--sites amino acids for each sequence.}
    \label{tab:seq}
\end{table}

\subsection{Local entropy of the basins}

Proteins are expected to tolerate random mutations in order to be evolutionary fit \citep{Guo2004}. Such tolerance can be characterized by the width of the neighborhood of a protein sequence $\bbsigma$, quantified by its local entropy difference $\Delta S_{T_s, \gamma}(\bbsigma)$ (Eq. \ref{eq:ds}), as done in the energy landscape of other kinds of complex systems \citep{Baldassi2016ent}.

\begin{figure}
    \centering
    \includegraphics[width=\linewidth]{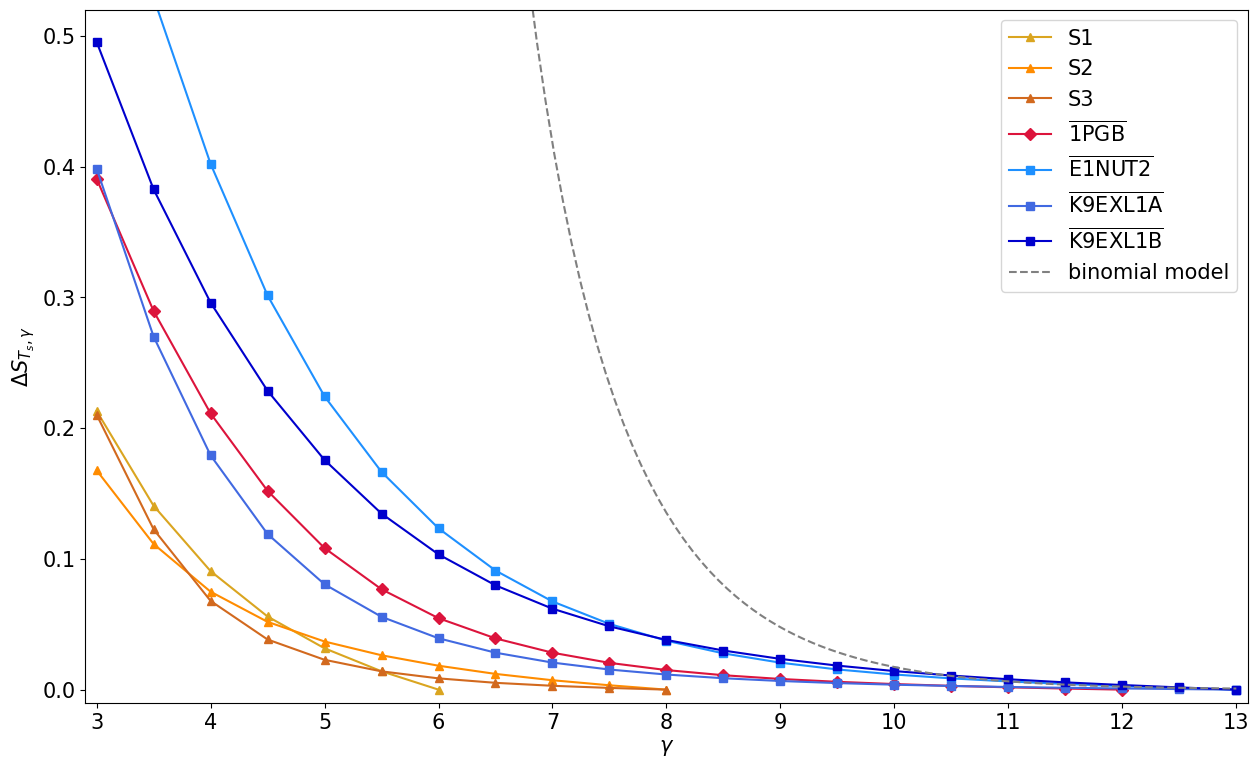}
    \caption{The local entropy difference $\Delta S_{T_s,\gamma}(\bbsigma)$ plotted as a function of the parameter $\gamma$ for (equilibrated) natural sequences and artificial ones. The grey dotted line represents the theoretical local entropy difference for the binomial model at $T_s = 3.2 \cdot 10^{-3}$ (see Table \ref{tab:binomial}).}
    \label{fig:sloc}
\end{figure}

We calculated the local entropy of the basins defined by natural sequences and by low--energy sequences sampled by the Monte Carlo algorithm but not present in nature (cf. Table \ref{tab:seq}). For each natural sequence (see Table \ref{tab:seq}), we first ran $\sim 10^4$ steps at $T_s=3.2\cdot 10^{-3}$ in order to obtain, for each basin, typical sequences for that temperature which still maintain the same K--sites of the starting natural ones. The sequences produced by this equilibration process, which is necessary to compare correctly the width of the basins within the framework of the canonical ensemble (cf. Fig. S10), are labeled with an overbar (cf. Fig. \ref{fig:sloc} and Table S1). We then proceeded to calculate the local entropy difference $\Delta S_{T_s,\gamma}$ as a function of the Lagrange multiplier $\gamma$ that controls the average distance from the representative sequence, using Eq. (\ref{eq:ds}).

Interestingly, at any value of $\gamma$ the local entropy of the basins defined by natural sequences is significantly larger than those of artificial sequences (see Fig. \ref{fig:sloc}). So, at any length scale around each $\bbsigma$, natural sequences display a wider energy basin than artificial ones, while the associated energies are similar (cf. Table S1).

The entropy $\Delta S_{T_s,\gamma}$ is for all folding sequences markedly lower than that of the binomial model (dashed curve in Fig.  \ref{fig:sloc}). This is not unexpected, as the manifold sampled at realistic $T_s$ is not convex, as would be if the binomial approximation were correct. 

Matching the dependence of $\Delta S_{T_s,\gamma}$ on $\gamma$ with that of $\overline{q}$ on $\gamma$, one can infer the dependence of $\Delta S_{T_s,\gamma}$ as a function of $\overline{q}$ (see Fig. S9 in the Supp. Mat.). This curve displays a linear growth, indicating that the number of low--energy sequences in the neighborhood of each $\bbsigma$ grows exponentially with the distance from it. 

\subsection{Searching for high--local entropy sequences}

A relevant question is then whether there is a way to find efficiently sequences in wide energy basins, avoiding those that lie in narrow minima. In the field of artificial neural network, this goal was achieved sampling the space of the network parameters with replicas whose mutual distances are coupled together by the Lagrange multiplier $\gamma$ and varying (annealing) slowly $\gamma$ until the system converges to a unique set of parameters \citep{Baldassi2016ent}. We have applied the same strategy to the space of protein sequences, as described in Sect. \ref{sect:replicas}.

\begin{figure}
    \centering
    \centerline{\includegraphics[width=1.1\linewidth]{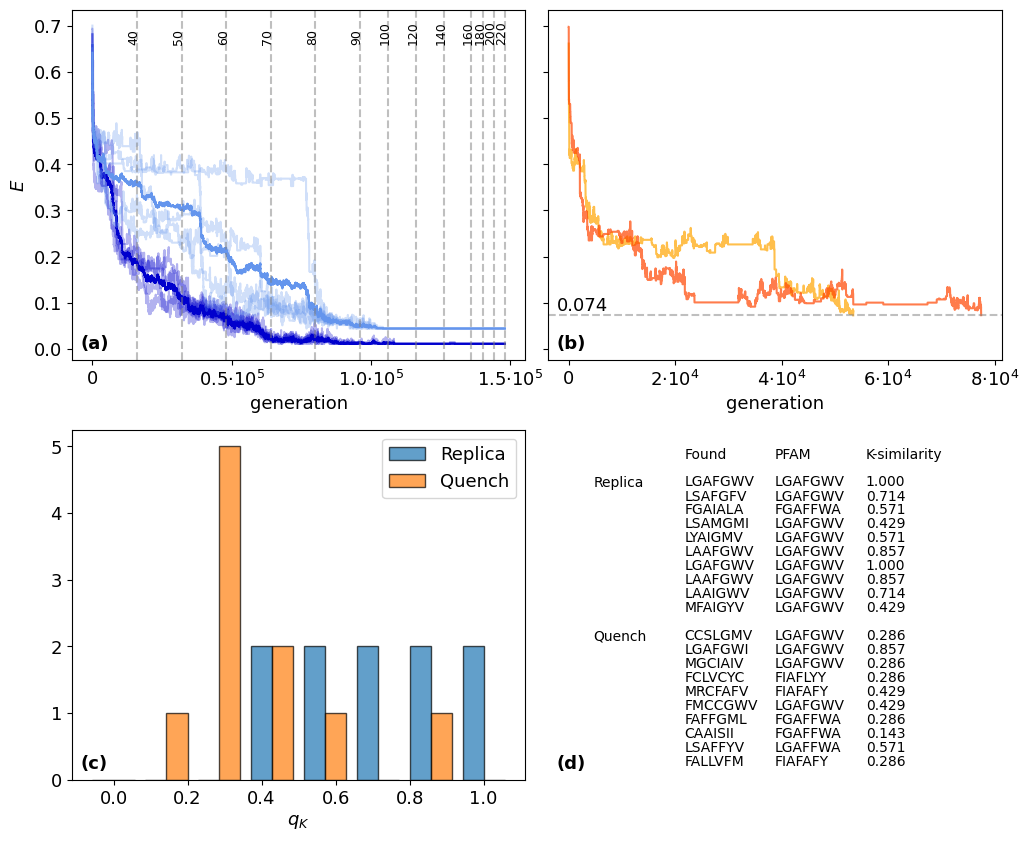}}
    \caption{
    (a) The effective energy of replicated simulations with $y = 5$. Transparent curves are the effective energy of each replica sequence, while the solid lines represent their mean. The vertical grey dotted lines indicate the value of $\gamma$ along the annealing.
    (b) The effective energy of quenched simulations. The horizontal grey dotted line is the mean effective energy at $T_s = 3.2 \cdot 10^{-3}$ (reported in the bottom left corner).
    (c) The non--normalized distributions of $q_K$ for the replicated (in blue) and quenched (in orange) simulations.
    (d) The K--sites found from the replicated and quenched simulations, compared to the most similar amino acids from the PFAM database and the associated $q_K$.
    }
    \label{fig:replicas}
\end{figure}

Starting from random sequences, the system can converge to a unique sequence of low energy with annealings of the order of $\sim 10^5$ steps (Fig. \ref{fig:replicas}a). We compared these sequences with those generated with quenches from infinite temperature to $T_s = 3.2 \cdot 10^{-3}$ (Fig. \ref{fig:replicas}b), recording a sequence when its energy reaches the average value at this temperature (cf. Fig. \ref{fig:energy}). 

In particular, we compared the K--sites of ten sequences obtained from the replica simulations with sequences obtained from ten temperature quenches. We defined $q_K$ as the maximum Hamming similarity between the K--sites of a simulated sequence with those of any natural sequence taken from the PFAM database. In this way, $q_K(\bbsigma)=1$ if there is at least a natural sequence displaying the same K--sites of the simulated sequence $\bbsigma$. The average $\overline{q_K}$ of the sequences obtained from the replica simulations is $0.71 \pm 0.21$, which is significantly larger than the value $0.39 \pm 0.20$ obtained from the quenches (cf. Fig. \ref{fig:replicas}c,d). Furthermore, the p--value (obtained from a t-test) for the two distributions is $2.4\cdot10^{-3}$. Thus, sequences in large basins are more similar from the point of view of K--sites combinations to natural sequences than other ones selected at random with comparable energy.

Molecular--dynamics simulations  of a sequence found by the replica algorithm (cf. Fig. S11 in the Sup. Mat) display a RMSD to the putative native structure of $0.14\pm 0.02$ nm, which is the more similar to the wild--type sequence 1pgb than those obtained from a temperature quench.

It is also worth mentioning that if sequences are let evolve for $\approx 3 \cdot 10^5$ generations after the quench, they can reach the widest basins where natural sequences lie (cf. Fig. S12 in the Sup. Mat.). This makes the case of protein sequences different from other complex systems for which the replica algorithm was originally developed, where the system is unable to reach wide basins with simple Monte Carlo moves \citep{Baldassi2016ent}. The main reason for this difference seems to be that the relevant dimensions of the sequence space are just the seven ones that define the K--sites. Thus, the effective dimension of this system is much smaller and it can therefore be sampled much more easily, than that of other complex systems.

We stress that the advantage of the replica algorithm to find wide basins is not much that of computational efficiency, that is marginal for a protein as small as protein G, but that of guaranteeing to ignore narrow basins, allowing us to distinguish the properties of wide minima (in the present case, the composition of K--sites) from those of minima at large. Our proof of concept however shows that the algorithm can be easily extended to the case of more complex, larger proteins.

\section{Discussion}

Characterizing the space of sequences folding to a well--defined native structure is useful to define the constraints that bind evolutionary trajectories of proteins. Recently developed machine--learning models like ESMFold are an efficient tool to define the landscape of foldable sequences. 

A concern intrinsically associated with this approach is the reliability of the predictions of the machine--learning model for sequences that are far from natural ones. A feature of ESMFold that makes it particularly suitable for our goal is that, at variance with other predictors, it does not use information from alignments of homologous sequences. Its prediction does not stem from what amino acids are observed in the very same sites in evolutionary--related proteins, but from the overall information coming from all available protein structures, which gives good predictions even for test sets not containing sequences homologous to those used for the training (cf. appendix B in ref. \cite{Hsu2022LearningStructures}). Consequently, its performance is not expected to drop as the sampling departs from the set of naturally--observed sequences. In fact, the molecular dynamics simulations we did starting from the predicted native conformation of sequences very far from the natural ones proved very stable. On the contrary, AlphaFold \citep{Jumper2021} did not produce any reasonable structure given the same sequences, since no homologous are found.

The picture that emerges from our sampling of the sequence space of protein G is that foldable sequences form a wide basin that contains all natural homologous and a constellation of smaller basins that display similar effective energy as the main one but that are narrower, displaying lower entropy. The different basins are characterized by different combinations of amino acids in a limited number of specific sites, here termed K-sites. The other amino acids, not belonging to K--sites, are rather free to mutate, thus generating a connected set of well--folding sequences up to very large Hamming distance from the wild--type. Only mutations in the K--sites seems to generate energy barriers corresponding to poorly--folding sequences. 

The presence in the sequence landscape of different basins characterized by specific choices of amino acids in few key sites of the protein was already found in minimal models \citep{Tiana2000} and in simplified models with knowledge-based potentials \citep{Tiana2009}. This fact suggests that it is not a consequence of the particular energy function used here, but it is a general feature of this kind of systems. Differently from what suggested in the case of simplified protein models, the key sites we found are not those critical for folding kinetics \citep{McCallister2000}.

An important result of this study is that natural sequences folding to the structure of protein G belong to a wide basin, which maximizes the local entropy. One could hypothesize that being able to accumulate several mutations while maintaining the same native structure is an evolutionary advantage for a protein.

Wide energy basins can be found very efficiently with algorithmic schemes borrowed from the theory of complex systems. These do not seem to mimic in any way the evolutionary dynamics of proteins but are indeed a fast computational tool. For a small protein like protein G, we saw that it is possible to find the widest basin simply with a Monte Carlo algorithm in a manageable computational time, even without resorting to local entropy minimization. However, it seems unlikely that the same can be done for larger system, in which the dimension of the sequence space is larger. On the other hand, the algorithm for entropy--driven search can be made more efficient than in the present proof of concept in a number of ways \citep{Pittorino2020}.

The structure of the energy landscape we found for protein sequences seems quite different from the typical landscapes of systems with complex interactions like spin glasses and constraint minimization problems, which are much more rugged and display a much larger number of well--separated basins. As a matter of fact, the ruling role that K--sites have on the effective energy of sequences makes their physical properties simpler than those of other complex systems.

Of course, foldability is just one of the constraints that the evolution of a protein must satisfy. Large language models can anyway be employed to define effective energies that encode other properties of sequences, like their thermodynamic stability or their binding properties to specific targets. The strategy developed here is agnostic of the physical meaning of the effective energy.


\section{Conclusions}

We defined an effective energy based on currently--available large language model and explored the energy landscape associated with the sequences folding to the native conformation of a small protein. This problem can be conveniently cast into the framework of the canonical ensemble of statistical physics, using the tools developed in this field. We found that folding sequences populate few basins of similar energy; one of them is much wider than the others and contain naturally--evolved sequences. Different basins are characterized by specific arrangement of the amino acids in a small subset of the sites of the protein. We showed that a computational algorithm based on replicated searchers can identify very efficiently the widest basins. 

\vspace{1cm}
\noindent Conflict of interest: the authors declare no conflict of interest.

\noindent The codes and the data can be downloaded freely from the   \href{https://dataverse.unimi.it/dataset.xhtml?persistentId=doi:10.13130/RD_UNIMI/WQ8QUA}{institutional repository} of the University of Milano.


\end{document}